# Safety and Security for Intelligent Infrastructure

Kevin Fu
University of Michigan

Ann Drobnis
Computing Community Consortium

Greg Morrisett
Cornell University

Elizabeth Mynatt
Georgia Tech
Computing Community Consortium

Shwetak Patel
University of Washington

Radha Poovendran
University of Washington

Ben Zorn[1]
Microsoft Research

**What is Intelligent Infrastructure?**

While our nation's cities and communities face increasingly serious challenges brought on by an aging and inadequate physical infrastructure, rapid advances in computing and communications have radically transformed cyberinfrastructure industries in the US economy. These technologies have the potential to similarly transform our cities and communities; improving efficiencies while increasing livability, improving resilience and disaster response thus saving lives and reducing economic destruction, and creating new economic opportunities while also reducing healthcare disparities and costs.[2]

**The Argument for Intelligent Infrastructure**

Increasingly, smart computing devices, with powerful sensors and internet connectivity, are being embedded into all new forms of infrastructure, from hospitals to roads to factories. These devices are part of the Internet of Things (IoT) and the economic value of their widespread deployment is estimated to be trillions of dollars, with billions of devices deployed. Consider the example of "smart meters" for electricity utilities. Because of clear economic benefits, including a reduction in the cost of reading meters, more precise information about outages and diagnostics, and increased benefits from predicting and balancing electric loads, such meters are already being rolled out across North America[3]. With residential solar collection, smart meters allow individuals to sell power back to the grid providing economic incentives for conservation. Similarly, smart water meters allow water conservation in a drought. Such infrastructure upgrades are infrequent (with smart meters expected to be in service for 20-30 years) but the benefits from the upgrade justify the significant cost. A long-term benefit of such upgrades is that unforeseen savings might be realized in the future when new analytic techniques are applied to the data that is collected. The same benefits accrue to any infrastructure that embeds increased sensing and actuation capabilities via IoT devices, including roads and traffic control, energy and water management in buildings, and public health monitoring.

**Potential Problems & Challenges**

While there are clear benefits to embedding sensing, computing, and communication into previously "dumb" infrastructure, this technology also creates new and challenging threats to safety, security, and privacy. These threats occur in multiple forms (from information leakage to petty cybercrime to cyberwar) and at multiple scales (issues with individual devices, issues with a single device networked to a server in the cloud, and emerging issues with

---

[1] With contributions from Nadya Bliss, Betsy Cooper and Jamie Winterton.
[2] Mynatt et al. (2017) "A National Research Agenda for Intelligent Infrastructure" CCC Led Whitepapers http://cra.org/ccc/resources/ccc-led-whitepapers/, last accessed April 12, 2017.
[3] For example, BC Hydro, in British Columbia has already installed 1.9 million smart meters.

large collections of devices coordinated for a specific purpose). These threats include:

- **Threats against individual devices**: Attacks against individual devices might include customers physically or electronically hacking their meter to cheat the power company or a cyber criminal turning off power in a house as a form of physical ransomware. Criminals might steal personal information being collected by a device to learn when an individual is away, or unscrupulous companies might steal personal data and resell it to the highest bidder.
- **Threats from the network**: Because smart devices are networked, the opportunity exists for attackers to interpose themselves between your device and the legitimate servers they communicate with, allowing attackers to steal valuable information or control the device through "man-in-the-middle" attacks that do not require them to compromise the device itself.
- **Emerging threats from collections of devices**: Because smart devices are deployed at scale, there are often hundreds of thousands of devices deployed and network connected running similar or identical hardware and software. Such deployments introduce new complexities for IT administrators to manage and represent a new emerging source of potential threat. In particular, well-resourced cyber criminals (such as nation states) might seek to compromise large numbers of devices simultaneously, potentially disrupting or destroying the power grid itself. Such scenarios are not hypothetical. In the Fall of 2016 cyber attackers used a botnet network of more than 100,000 compromised smart cameras to launch a distributed denial of service (DDoS) against the Internet backbone[4].

While these examples focus on smart meters and the power grid, these threats generalize to other traditional forms of infrastructure as well. Devices embedded in roads can be hacked to disrupt transportation. Devices embedded in buildings might be hacked to cause physical damage or allow security breaches. Agriculture sabotage could threaten a competitor's crops. Perhaps more troubling, threats to health-focused sensors and interventions could aim to cause direct physical harm.

**Research Investments in Security Solutions**
Addressing the threats highlighted in the previous section will require coordinated and timely research investments that extends beyond current cybersecurity research. Here we discuss some of the broad themes that research will encompass as well as research areas to address in the specific categories of threats mentioned above. Several these concerns are also discussed at length in the Report from the President's Commission on Enhancing National Cybersecurity published in December 2016.

**Theme: Protecting Individual Devices**
A significant amount of academic and industrial research has been invested in understanding the challenges of securing individual devices. Beyond testing, researchers have demonstrated formal verification approaches that ensure device software and hardware conforms to a mathematical specification. Programs such as DARPA HACMS, where drone software created using formal methods was shown to be unattackable, are showing promise. While progress has been made, it's clear from global headlines that improvements in individual device security - including those that ensure a user's privacy - are still sorely needed. These topics should be at the center of a large-scale research agenda in cybersecurity.

Some of the qualities of devices installed in Intelligent Infrastructure are unique and present important opportunities for research investment. For example, such devices often have one or more physical sensors. Leveraging an understanding of the physics of sensors and the potential for multi-sensor cross-checking allows such devices to be tested for correctness in new ways. While verifying an entire general purpose computer (running Linux for example) may be very difficult and expensive, relatively simple devices embedded in infrastructure might be simple enough that formal verification methods are both possible and economical. Overall, the dollar cost of applying verification methods to larger software systems needs to be greatly reduced from what it costs now.

---

[4] https://www.theguardian.com/technology/2016/oct/26/ddos-attack-dyn-mirai-botnet

**Theme: Device Life Cycle and Graceful Degradation**
Given that devices installed in infrastructure need to last for decades, several top-level concerns arise related to maintaining the hardware and software in the devices over their entire life cycle. In particular, software updates will be necessary to upgrade the capabilities of the devices as well as patch bugs, etc. Many existing smart devices do not have any capability to be updated. Moreover, because updates can be used by attackers as a means of corrupting a device, creating a secure channel for updating devices is a key requirement for any deployment. Ensuring such a secure channel, especially for devices with limited resources, is an area of active research. Beyond updates, the requirements for extended support of devices over their entire life cycle must be considered. If the company selling the device goes out of business, what is the impact on continued maintenance of the device? So-called "zombie devices" that still exist but are no longer maintained must be accounted for. If devices fail or degrade, how does it impact the overall infrastructure? Approaches that allow for graceful degradation of the infrastructure if specific technology components fail are required.

**Theme: Protecting the Client, Server, and Network**
Even if the individual device is secure, the potential for stealing information or misdirecting the device from the network is possible. Two approaches can mitigate such challenges and warrant further research. First, establishing a strong understanding of identity using cryptography can allow two parties (e.g., client and server) to interact with greater confidence. Increasingly, cryptography is also being used within an application to ensure that the application executes correctly and does not reveal any private information to outsiders even if the host operating system is compromised. In addition, techniques that allow verified computation on the device (where data is never sent over the network) can prevent intruders from eavesdropping on private information. Finally, to protect individual privacy overall, techniques such as differential privacy provide mathematical guarantees that limit the amount of information a third party can learn about an individual by analyzing collected data.

**Theme: System Interdependencies, Compartments, and Points of Failure**
In general, as smart technology creates greater interdependencies (for example between the HVAC system and the lighting system in a building), the potential arises for unforeseen circumstances when one system fails to operate as expected. Failures in one component that another depends on can propagate and cause even bigger problems. Cascading failures occur in practice and have resulted in major disruptions in data centers. Just as ships have been carefully engineered with multiple watertight compartments, Intelligent Infrastructure systems need well-defined compartments that prevent a catastrophic failure due to a single error. Moreover, given that infrastructure can be a target of terrorist attacks, the resiliency of the infrastructure to multiple targeted failures must be understood.

**Theme: Protecting Collections of Devices**
An emerging threat that arises when tens of thousands of similar devices are deployed in infrastructure is the potential for an attacker to compromise many related devices at the same time, giving them greater potential for damage. If the protections on individual devices fail, then adequate failsafe mechanisms are required to understand that such an attack has occurred and take reactive measures. This challenge also relates to the need to understand interdependencies. What other devices depend on these compromised systems? How will, say, shutting off or rebooting these devices affect other services that might depend on them? Managing a collection of devices is even harder than managing an individual device and we know that much vulnerability in individual devices occur because of simple configuration errors such as weak passwords, unnecessary entry points, etc. New techniques are required to reduce the burden of understanding the current state of large collections of devices and manage them as a unit. For example, all security analysis requires understanding the "threat model", that is, the assumptions made about what the attacker can and cannot do. Because the threats related to large device collections are relatively new, how they might be harnessed by an attacker is less understood, resulting in more likely unforeseen consequences.

**Theme: Humans are Essential and Education and Simplicity are Necessary**
A key principle in operating services that run 24/7 is that there is always a human operator. This individual is necessary because failures occur and complexities arise that are impossible for any software to anticipate. Similarly, with Intelligent Infrastructure, individuals will be necessary to continuously oversee the correct and secure operation of the infrastructure. Human error is also a cause for many failures in data center operations because understanding the complexities of such systems is challenging. As a result, research is needed in finding ways to make a human operator most effective, including giving them tools to be well-informed as to the state of the system or the potential impact of actions they take. Likewise, research that can reduce complexity and improve the ability of a human to make well-informed decisions is valuable. Another strategy to improve human oversight is to recognize that humans working in teams can be more effective than individuals and to invest in research and training to make teams of individuals more effective. Finally, given the widespread deployment of infrastructure, workforce development that trains individuals and teams to oversee these systems will be necessary.

Beyond educating the operators, it is also important that individuals using and affected by technology deployed as part of Intelligent Infrastructure understand the implications of the technology and how it impacts them. For example, in the case of smart meters, owners should be aware of the privacy implications of the devices as well as what precautions they should take regarding protecting the device both physically and against cybersecurity attacks (for example to prevent someone from hacking their account).

**Theme: Engineers and Computer Scientists Need Join Forces**
While much of the discussion so far has focused on security, cyber-physical systems introduce the possibility that any security vulnerability might be transformed into a safety hazard. Most computer science security experts do not have significant training in the engineering skills required to build devices that are safer. Likewise, many engineers, including mechanical and electrical engineers, do not automatically think about how a hacker might exploit their designs to harm people or cause physical damage (so-called adversarial thinking). Research that bridges this divide is necessary as society becomes increasingly dependent on cyber-physical systems. For example, finding ways that sensors can be leveraged to improve the overall security and safety of such systems would be valuable.

**Summary**
We have highlighted why next-generation infrastructure will necessarily include embedded smart devices that improve the efficiency, quality, and value of the modern infrastructure. Such devices change the way in which we think about infrastructure and many of the challenges already understood about computer security in general apply to these systems. Security in infrastructure is essential because with cyber-physical systems, security vulnerabilities can translate directly to public safety hazards. We have discussed how such systems require expansive thinking about safety and security and we have described the kinds of research that will improve these qualities.

Beyond the technical investments discussed, we believe that deep investments in workforce training are required. Well-informed operators using the most effective analysis tools are critical to maintaining Intelligent Infrastructure safety and security. In addition, because much of future infrastructure will be created by the private sector, it is essential for the federal government to cooperate with the private sector in encouraging the use of "best practices" regarding the design and implementation of such systems. Beyond sharing best practices, which is an ongoing effort, finding creative ways to incentivize the private sector to build new infrastructure to the highest safety and security standards is important. Only with high public confidence in its safety, security, and privacy will the full potential of Intelligent Infrastructure be realized.

*This material is based upon work supported by the National Science Foundation under Grant No. 1136993. Any opinions, findings, and conclusions or recommendations expressed in this material are those of the authors and do not necessarily reflect the views of the National Science Foundation.*